\documentclass[aps,prl,letterpaper,superscriptaddress,twocolumn]{revtex4}

\usepackage{amstext,amsmath,amssymb,ulem,amsthm}
\usepackage{color}

\usepackage{times}
\usepackage{algorithm}
\floatname{algorithm}{Protocol}
\usepackage{ftnxtra}

\usepackage[T1]{fontenc}
\usepackage[utf8]{inputenc}
\usepackage{graphicx}
\usepackage[colorlinks]{hyperref}

\theoremstyle{plain}

\theoremstyle{definition}

\newcommand{\beq}{\begin{equation}}
\newcommand{\eeq}{\end{equation}}

\newcommand{\ket} [1] {\vert #1 \rangle}
\newcommand{\bra} [1] {\langle #1 \vert}

\newcommand{\ba}{\begin{align}}
\newcommand{\ea}{\end{align}}
\newcommand{\bea}{\begin{eqnarray}}
\newcommand{\eea}{\end{eqnarray}}

\usepackage{tikz}	
\usepackage[hang,small,bf]{caption}  
\usetikzlibrary{backgrounds,fit,decorations.pathreplacing}  
\makeatletter

\@ifundefined{textcolor}{}
{%
 \definecolor{BLACK}{gray}{0}
 \definecolor{WHITE}{gray}{1}
 \definecolor{RED}{rgb}{1,0,0}
 \definecolor{GREEN}{rgb}{0,.4,0}
 \definecolor{BLUE}{rgb}{0,0,1}
 \definecolor{CYAN}{cmyk}{1,0,0,0}
 \definecolor{MAGENTA}{cmyk}{0,1,0,0}
 \definecolor{YELLOW}{cmyk}{0,0,1,0}
 }

\makeatother

\usepackage{bm}
\usepackage{mathtools}

\def\id{I}

\def\1{\mat{\id}}
\def\mat#1{\mathbf{#1}}

\renewcommand{\sout}[1]{}

\begin{document} 

\title{{Quantum assisted Gaussian process regression}}
\author{Zhikuan Zhao}
\affiliation{Singapore University of Technology and Design, 8 Somapah Road, Singapore 487372}
\author{Jack K. Fitzsimons} 
\affiliation{Department of Engineering Science, University of Oxford, Oxford OX1 3PJ, UK}
\author{Joseph F. Fitzsimons}
\email{joseph_fitzsimons@sutd.edu.sg}
\affiliation{Singapore University of Technology and Design, 8 Somapah Road, Singapore 487372}
\affiliation{Centre for Quantum Technologies, National University of Singapore, 3 Science Drive 2, Singapore 117543}
\begin{abstract}
Gaussian processes (GP) are a widely used model for regression problems in supervised machine learning. Implementation of GP regression typically requires $O(n^3)$ logic gates. We show that the quantum linear systems algorithm [Harrow et al., Phys. Rev. Lett. 103, 150502 (2009)] can be applied to Gaussian process regression (GPR), leading to an exponential reduction in computation time in some instances. We show that even in some cases not ideally suited to the quantum linear systems algorithm, a polynomial increase in efficiency still occurs.
\end{abstract}

\maketitle
\date{today}
\maketitle
Gaussian processes (GP) are commonly used as powerful models for regression problems in the field of supervised machine learning, and have been widely applied across a broad spectrum of applications, ranging from robotics, data mining, geophysics (where they are referred to as kriging) and climate modelling all the way to predicting price behaviour of commodities in financial markets. Although GP models are becoming increasingly popular in the community of machine learning, it is known to be
computationally expensive, hindering their widespread adoption. A practical implementation of Gaussian process regression (GPR) model with $n$ training points typically requires $O(n^3)$ basic operations \cite{GP}. This has lead to significant effort aimed at reducing the computational cost of working with such models, with investigations into low rank approximations of GPs \cite{quinonero2005unifying}, variational approximations \cite{hensman2013gaussian} and Bayesian model combination for distributed GPs \cite{deisenroth2015distributed}. 

The enterprise of designing quantum algorithms has come a long way since Feynman's original vision of utilizing the exponential Hilbert space in quantum mechanics to simulate quantum physics \cite{Feynman}. Among the most celebrated results are Shor’s factoring algorithm \cite{Shor} and Grover's search algorithm \cite{Grover}. More recently, machine learning has emerged as a field in which quantum algorithms can have a dramatic impact \cite{aimeur2006machine,pudenz2013quantum,QML, QPC, superunsuper, QVSM}. Of particular interest to the field of machine learning, in 2009 Harrow \textit{et} \textit{al} presented a quantum algorithm which produces a superposition state $\ket{x}$ with $\epsilon$ error, such that the vector $\ket{x}$ solves the linear system $A\ket{x}=\ket{b}$ \cite{HHL}. For an $n\times n$ $s$-sparse matrix $A$ with condition number (the ratio between its largest and smallest eigenvalues) $\kappa$, the runtime roughly grows as $\tilde{O}(\log(n)\kappa^2{s}^2/\epsilon)$ (where $\tilde{O}$ suppresses slower growing contribution), while the classical counter-part, matrix inversion algorithms for sparse matrices runs at $O(n\kappa{s}\log(1/\epsilon))$ using the conjugate gradient method \cite{CMM}.

In this work we demonstrate that the Quantum linear algorithm (QLA) described in \cite{HHL} can be used to dramatically speed up computation in GPR. We start with reviewing both the basics of classical GPR and its conventional implementation, as well as the original QLA. We then propose a procedure of applying QLA to GPR modelling, and discuss the performance of such a procedure. Finally, we address potential caveats of the procedure, and discuss potential issues arising in the application of our procedure to specific GPR problems. 

Supervised machine learning endeavours to learn the relationship between the input and output of a system based on a set of examples, referred to as the observations of a training set. Gaussian processes offer a number desirable properties in doing this such as ease in expressing uncertainty, the ability to model a wide range of behaviours under a simple parametrisation, and admitting a natural Bayesian interpretation.

Given a training set $\mathcal{T}=\{\mathbf{x}_i,y_i\}^{n-1}_{i=0}$, containing $n$ $d$-dimensional inputs, $\{\mathbf{x}_i\}_{i=0}^{n-1}$, and corresponding outputs, $\{y_i\}_{i=0}^{n-1}$, we wish to model the latent function $f(\mathbf{x})$
\begin{align}
 y = {f}(\mathbf{x}) + \epsilon_{\text{noise}},
\end{align}
where $\epsilon_{noise}\sim\mathcal{N}(0,\sigma_n^2)$ is independent and identically distributed Gaussian noise. As such, given a new input (we will call it a ``test point''), $\mathbf{x}_*$, we aim to have a predictive distribution of ${f}_*={f}(\mathbf{x}_*)$.

A GP approach to such a regression problem models the behaviour of latent variables $\{{f}(\mathbf{x_i})\}_{i=0}^{n-1}$ as a joint multi-dimensional Gaussian distribution \cite{GP}. A GPR model is fully specified by a mean function $m(\mathbf{x})=\mathbb{E}[{f}(\mathbf{x})]$ and a covariance function (also known as kernel) $k(\mathbf{x},\mathbf{x}^\prime)=\mathbb{E}[({f}(\mathbf{x})-m(\mathbf{x}))({f}(\mathbf{x}^\prime)-m(\mathbf{x}^\prime))]$, where $\mathbb{E}[z]$ denotes the expectation value of $z$. Without loss of generality we can assume the GPR to have zero prior mean. We write 
\begin{align}
 {f}(\mathbf{x})\sim\mathcal{G}\mathcal{P}(m(\mathbf{x}),k(\mathbf{x},\mathbf{x}^\prime)).
\end{align}
Conditioning on the training data set, we write the predictive distribution of ${f}_*$ in the form of a multi-variable Gaussian distribution \cite{gaussian1998},
\begin{align}
p({f}_*|\mathbf{x}_*,\mathcal{T})\sim\mathcal{N}(\bar{{f}_*},\mathbb{V}[{f}_*]) .
\end{align} 
The central goal in GPR models is to predict the mean of this distribution, also known as the linear predictor $\bar{{f}_*}$, and its variance $\mathbb{V}[{f}_*]$ given the inputs $\{\mathbf{x}_i\}$, observed output vector \textbf{y}, the covariance functions $k$, and the noise variance $\sigma_n^2$. For simplicity, we consider only one test point, although the same principle applies to an array of test points. Let the entries of the vector $\textbf{k}_*$ denotes the covariance functions between the test point $\mathbf{x}_*$ and each of the $n$ input points in the training set, such that  $\textbf{k}_*=\mathbb{E}[({f}(\mathbf{x})-m(\mathbf{x}))({f}(\mathbf{x}_*)-m(\mathbf{x}_*))]$. We denote by $K$ the $n\times n$ matrix of covariance functions between the input points in the training set. Following the derivation presented in \cite{GP}, we obtain the moments of a zero mean GP as
\begin{align}
 \bar{{f}_*}&=\textbf{k}_*^T(K+\sigma_n^2{I})^{-1}\textbf{y} \label{eq:mean}\\
 \mathbb{V}[{f}_*]&=k\left(\textbf{x}_*,\textbf{x}_*\right)-\textbf{k}_*^T(K+\sigma_n^2{I})^{-1}\textbf{k}_*.\label{eq:variance} 
\end{align}

We will outline a typical implementation of GPR on a classical machine: The first step is to compute the Cholesky decomposition of $(K+\sigma_n^2{I})$, that is, to find the lower-triangular matrix $L$, known as the Cholesky factor, such that $(K+\sigma_n^2{I})=LL^T$. The computation of the Cholesky factor is known to be numerically stable, and has runtime proportional to $n^3$. Writing $\boldsymbol{\alpha}=(K+\sigma_n^2{I})^{-1}\textbf{y}$, the linear predictor is rewritten as $\bar{{f}_*}=\textbf{k}_*^T\boldsymbol{\alpha}$. Computing $\mathbf{\alpha}$ then amounts to solving $LL^T\boldsymbol{\alpha}=\textbf{y}$. Using the backslash notation $L\backslash \textbf{y}$ to denote the vector $\textbf{y}^\prime$ which solves the triangular linear system $L\textbf{y}^{\prime}=\textbf{y}$, $\boldsymbol{\alpha}=L^T\backslash L\backslash\textbf{y}$ can be computed by solving two triangular systems, taking time proportional to $n^2$. Similarly,  $\mathbb{V}[{f}_*]=k(\mathbf{x}_*,\mathbf{x}_*)-(L\backslash\textbf{k}_*)^T(L\backslash\textbf{k}_*)$ can be computed in with a number of basic arithmetic operations proportional to $n^2$. Therefore the total runtime of computing the mean and variance of GPR amounts to $O(n^3)$. For problems involving thousands of input points, the exact inference in GPR become intractable, which motivates the search for a quantum approach to speed-up this computation. This is where the QLA, first introduced in \cite{HHL}, offers an advantage.

We now give an outline of the original QLA to solve the linear system, ${A}\mathbf{x}=\mathbf{b}$:
\begin{itemize}
\item Prepare the state $\ket{\mathbf{b}}=(\mathbf{b}^T\mathbf{b})^{-1/2}\sum\limits_{i=0}^{n-1}b_i\ket{i}$ to encode the vector $\mathbf{b}$. Prepare an ancilla register in a supersition state $\frac{1}{\sqrt{T}}\sum_{\tau = 0}^T \ket{\tau}$.
\item Simulate ${A}$ as a Hamiltonian at time $\tau$ applied to $\ket{\mathbf{b}}$ using phase estimation techniques described in \cite{pe}, and expand $\ket{\mathbf{b}}$ into the eigenbasis of ${A}$. We obtain the state after evolution:
\begin{align}
\ket{\phi_1}=\frac{1}{\sqrt{T}}\sum\limits_{i=0}^{n-1}\sum\limits_{\tau=0}^{T-1}\ket{\tau}e^{i\lambda_it_0\tau/T}\beta_i\ket{\mu_i},
\end{align} 
where $\lambda_i$ are the eigenvalues and $\ket{\mu_i}$ are the eigenvectors of $A$ . Each $\ket{\mu_i}$ is associated a complex probability amplitude $\beta_i$. The time $t_0$ here scales linearly with $\kappa$. The time period in the second sum is chosen to be some large $T$ as in the improved phase-estimation procedure described in \cite{clock}.
\item Apply the quantum Fourier transform (QFT) \cite{NC} to the first register in $\ket{\phi_1}$ and obtain
\begin{align}
\ket{\phi_2}=\sum\limits_{i=0}^{n-1}\beta_i\ket{\tilde{\lambda}_i}\ket{\mu_i}.
\end{align}
\item Introduce an ancilla qubit and perform a controlled rotation on it to yield the extended state
\begin{align}
\ket{\phi_3}=\sum\limits_{i=0}^{n-1}\beta_i\ket{\tilde{\lambda}_i}\ket{\mu_i}\left(\sqrt{1-\frac{c^2}{\lambda_i^2}}\ket{0}+\frac{c}{\lambda_i}\ket{1}\right).
\end{align}
Here the constant $c$ is used to ensure that the rotation is bounded by $\pi$. 
\item Finally, reverse the phase estimation to uncompute $\ket{\tilde{\lambda}_i}$. Measure the ancilla qubit. A result of $\ket{1}$ result in the state vector encoding the solution of ${A}\mathbf{x}=\mathbf{b}$,
\begin{align}
\ket{\mathbf{x}}=\ket{\phi_{final}}=\sum\limits_{i=0}^{n-1}\frac{\beta_i}{\lambda_i}\ket{\mu_i}
\end{align}
\end{itemize}
For simplicity, we have omitted global normalization factors in the last step of the above outline, and have assumed ${A}$ to be Hermitian throughout. However, \cite{HHL} also includes a treatment to ``Hermitianize'' a general ${A}$, which involves building an anti-diagonal block matrix with the elements of $A^\dagger$ and $A$ in the lower and upper half of the new matrix respectively. Once $\ket{\mathbf{x}}$ has been produced, quantum measurements  can be used to estimate expectation values corresponding to some desired quantity of the form $\bra{{x}}{M}\ket{{x}}$. For a quick account of the runtime, we note that the $\epsilon$-error runtime in phase estimation scales quadratically with the sparsity $s$ of $A$ \cite{pe}, and $t_0=O(\kappa/\epsilon)$, the repetition needed to obtain the desired measurement on the ancilla qubit scales proportionally to $\kappa$, and hence the total runtime amounts to $\tilde{O}(\log(n)\kappa^2{s}^2/\epsilon)$. We direct interested readers to \cite{HHL} and its supplementary material for a detailed error and runtime analysis.

Soon after the original QLA was proposed, Clader  \textit{et} \textit{al} extended the algorithm to include an efficient method to prepare the input encoding using entangled states in $O(1)$ query complexity with the help of an oracle which calculates the amplitude and phase components of the vector $\ket{\mathbf{b}}$ \cite{precondi}. In the same paper, the authors also developed a scheme to precondition ${A}$ taking $O(s^3)$ runtime overhead to suppress the growth of $\kappa$. A very recent result by Childs \textit{et} \textit{al} further modified the QLA into a novel algorithm based on implementing operators with Fourier or Chebyshev series representation, which further suppressed the runtime through a logarithmic $\epsilon$-precision dependence \cite{Childs2015}. The sparse-dependent efficiency of the Hamiltonian simulation stage in QLA was also subsequently optimized by an algorithm revealed in \cite{berry}, which runs almost linearly in $s$, improving the $s$-dependence of the original QLA runtime to $O(\frac{s\kappa^2}{\epsilon}\text{polylog}(s\kappa/\epsilon)\text{poly}(\log(n)))$ \cite{Childs2015}. Absorbing the slower growing terms into $\tilde{O}$, the improved runtime amounts to $\tilde{O}(\log(n)\kappa^2{s}/\epsilon)$.

Despite the promising exponential speed-up the quantum linear algorithm (QLA) can potentially provide, one has to apply it with care. As Aaronson accurately described in \cite{Aaronson}, there are four practical aspects that needs particular care in any application of the original QLA: (1) The time taken to prepare $\ket{\mathbf{b}}$ encoding $\mathbf{b}$ needs to be taken into account; (2) the matrix ${A}$ has to be robustly invertible, $\kappa$ needs to grow at most polylogarithmically in $n$ to maintain an exponential speed-up; (3) one also needs to address the sparseness contribution to the total runtime, since the general phase estimation sub-routine in QLA costs polynomial time in $s$; (4) Although the output of QLA is the state $\ket{\mathbf{x}}$, there is no efficient way to extract entries of the vector $\mathbf{x}$. One needs to make sure that the matter of practical interest does not span the full glory of $\mathbf{x}$, but is restricted only to information which is accessible with relatively few copies of $\ket{\mathbf{x}}$. For example, one can efficiently estimate quantities such as $\bra{\mathbf{x}}{M}\ket{\mathbf{x}}$, where ${M}$ is some Hermitian matrix of interest which can be efficiently implemented as an observable, since this simply amounts to the expectation value of the observable $M$ on $\ket{\mathbf{x}}$. We now introduce a procedure for applying QLA to Gaussian process regression, and then address each of these practicality concerns.
 
We observe from Equations \ref{eq:mean} and \ref{eq:variance} that the computation of ${f}_*$ and $\mathbb{V}[{f}_*]$ involves solving linear systems of the forms $(K+\sigma_n^2{I})\boldsymbol{\alpha}=\mathbf{y}$ and $(K+\sigma_n^2{I})\boldsymbol{\eta}=\mathbf{k}_*$ respectively, where $\mathbf{k}_*^T\boldsymbol{\alpha}={\bar{f}}_*$ and $k\left(\textbf{x}_*,\textbf{x}_*\right)-\mathbf{k}_*^T\boldsymbol{\eta} =\mathbb{V}[{f}_*]$. The common linear structure suggests we can apply QLA to extract useful information.

In order to compute these expectations we will extend the quantum linear systems algorithm in two ways. First, we need an efficient method to prepare a state $\ket{\mathbf{v}}$ from a classical representation of a vector $\mathbf{v}$ of length $n$. To achieve this we use an approach based on quantum random access memory (QRAM) \cite{RAM} introduced in \cite{precondi}, which we modify to allow preparation of sparse (or approximately sparse) vectors. To prepare a state corresponding to the $s_\mathbf{v}$-sparse vector $\mathbf{v}$ with entries $v_i$, a register is prepared in a superposition $s_\mathbf{v}^{-1/2}\sum_{i: v_i \neq 0} \ket{i} \otimes \ket{0}$. Using the index stored in the first register, the ancilla register is rotated based on the $i$th entry of $\mathbf{v}$ such that the state of the system is \[ \ket{\tilde{\mathbf{v}}} = \frac{1}{\sqrt{s_\mathbf{v}}}\sum_{i:v_i\neq 0}^n \ket{i}\otimes (\sqrt{1-c_\mathbf{v}^2v_i^2}\ket{0} + c_\mathbf{v} v_i\ket{1}),\]
where $c_\mathbf{v} \leq \min_i |v_i|^{-1}$. Conditioned on the ancilla qubit being in state $\ket{1}$, the first register is in state $\ket{\mathbf{v}} = \frac{\mathbf{v}}{||\mathbf{v}||}$.

The second element necessary is a mechanism to estimate $\langle \mathbf{u} | \mathbf{v} \rangle$ for a given pair of real vectors $\mathbf{u}$ and $\mathbf{v}$. While the square of this value is accessible via a controlled-swap test, as discussed in \cite{Liming}, we require information about both the magnitude and sign of this inner product which is not accessible with such a test. In order to estimate the inner product, we instead apply a modified version of the state preparation procedure, where an additional ancilla qubit initially prepared in state $\frac{1}{\sqrt{2}}(\ket{0}+\ket{1})$ is used to determine whether the target state is $\ket{\mathbf{u}}$ or $\ket{\mathbf{v}}$. This results in a joint state 
 \begin{align}
 \ket{\Phi_{\mathbf{u},\mathbf{v}}} =& \frac{1}{\sqrt{2s_\mathbf{u}}}\sum_{i:u_i \neq 0} \ket{0}\ket{i} \left(\sqrt{1-c_\mathbf{u}^2 u_i^2}\ket{0} + c_\mathbf{u}u_i \ket{1}\right)\nonumber\\
 &+ \frac{1}{\sqrt{2s_\mathbf{v}}}\sum_{i:v_i \neq 0} \ket{1} \ket{i} \left(\sqrt{1-c_\mathbf{v}^2 v_i^2}\ket{0} + c_\mathbf{v}v_i \ket{1}\right). \nonumber
\end{align}
Then the expectation value of the operator $X \otimes I \otimes \ket{1}\bra{1}$ is $s_\mathbf{u}^{-1/2}s_\mathbf{v}^{-1/2}c_\mathbf{u} c_\mathbf{v} \mathbf{u}^T \mathbf{v}$.

These two elements can be combined with the quantum linear systems algorithm to compute quantities of the form $\mathbf{u}^T A^{-1} \mathbf{v}$ as follows:
\begin{enumerate}
\item Initialize the system in state $\ket{+}_A\ket{0}_B\ket{0}_C\ket{0}_D$, where $\ket{+}=\frac{1}{\sqrt{2}}(\ket{0} + \ket{1})$ and where $A$, $B$, $C$ and $D$ label distinct registers.
\item Conditioned on register $A$ being in state $\ket{0}$, prepare registers $B$ and $C$ in state $\ket{\tilde{\mathbf{u}}}$ such that the ancilla qubit is placed in register $C$ with the remainder of the state in register $B$, and apply $X$ to register $D$.
\item Conditioned on register $A$ being in state $\ket{1}$, prepare registers $B$ and $C$ in state $\ket{\tilde{\mathbf{v}}}$ such that the ancilla qubit is placed in register $C$ with the remainder of the state in register $B$. 
\item Conditioned on both registers $A$ and $C$ being in state $\ket{1}$, apply the quantum linear systems algorithm to register $B$ using register $D$ as the ancilla used in the QLA. A fifth register $E$ is used for the phase estimation step in the QLA, but since it is returned to the zero state, we do not explicitly include it in the description of the states after each step.
\item Measure the observable $M=X_A I_B \ket{1}\bra{1}_C \ket{1}\bra{1}_D$.
\end{enumerate}
The measurement result is then a Bernoulli random variable with expectation value $c s_\mathbf{u}^{-1/2}s_\mathbf{v}^{-1/2} c_\mathbf{u} c_\mathbf{v} \mathbf{u}^T A^{-1} \mathbf{v}$. To see this, note that the state of the system after the fourth step is given by
\begin{align}
&\frac{1}{\sqrt{2s_\mathbf{u}}}\ket{0}_A\sum_{i:u_i\neq 0} \ket{i}_B \left(\sqrt{1-c_\mathbf{u}^2u_i^2}\ket{0}_C + c_\mathbf{u} u_i\ket{1}_C\right)\ket{1}_D\nonumber\\
&+\frac{1}{\sqrt{2s_\mathbf{v}}} \ket{1}_A\sum_{i:v_i\neq 0} c_\mathbf{v} \beta_i \ket{\mu_i}_B \ket{1}_C \left(\sqrt{1-\frac{c^2}{\lambda_i^2}}\ket{0}_D + \frac{c}{\lambda_i}\ket{1}_D\right)\nonumber \\
&+\frac{1}{\sqrt{2s_\mathbf{v}}} \ket{1}_A\sum_{i:v_i\neq 0}  \sqrt{1-c_\mathbf{v}^2v_i^2}  \ket{i}_B \ket{0}_C \ket{0}_D\nonumber
\end{align}
In the above $\ket{\mu_i}$ is taken to be the eigenvector of $A$ corresponding to eigenvalue $\lambda_i$, and $\{\beta_i\}$ are taken to be the coordinates of $\mathbf{v}$ in the basis $\{\ket{\mu_i}\}$. Projecting this state onto $\ket{1}$ for registers $C$ and $D$ leads to the subnormalized state
\begin{align}
\frac{1}{\sqrt{2s_\mathbf{u}}}\ket{0}_A\sum_{i:u_i \neq 0} c_\mathbf{u} \gamma_i\ket{\mu_i}_B+\frac{1}{\sqrt{2s_\mathbf{v}}}\ket{1}_A\sum_{i:v_i \neq 0}^n \frac{c}{\lambda_i} c_\mathbf{v} \beta_i \ket{\mu_i}_B,\nonumber
\end{align}
where $\{\gamma_i\}$ are taken to be the coordinates of $\mathbf{u}$ in the basis given by $\{\ket{\mu_i}\}$.
Thus, the expectation value for the measurement in the final step is
\[
\sum_i \frac{1}{4}\left(\left(\frac{c_\mathbf{u}}{\sqrt{s_\mathbf{u}}} \gamma_i + \frac{c_\mathbf{v}c}{\sqrt{s_\mathbf{v}}} \frac{\beta_i}{\lambda_i}\right)^2-\left(\frac{c_\mathbf{u}}{\sqrt{s_\mathbf{u}}} \gamma_i - \frac{c_\mathbf{v}c}{\sqrt{s_\mathbf{v}}} \frac{\beta_i}{\lambda_i}\right)^2\right).
\] Since this is equal to $\frac{c_\mathbf{u}c_\mathbf{v}c}{\sqrt{s_\mathbf{u}s_\mathbf{v}}} \mathbf{u}^T A^{-1}\mathbf{v}$, the expectation value for the measurement in the final step must match this value.

The algorithm outlined above can be used to construct a quantum algorithm for approximating both the linear predictor and variance in GP regression, as follows:
\begin{itemize}
\item To approximate the linear predictor, take $\mathbf{u} = \mathbf{k}_*$, $A = K-\sigma_n^2I$ and $\mathbf{v}=\mathbf{y}$. Since $K$ is positive semi-definite (or approximately so) the minimum eigenvalue of $K-\sigma_n^2 I$ is at least $\sigma_n^2$, and hence we take $c=\sigma_n^2$ in each run of the QLA. This yields $\langle M\rangle = \frac{\sigma_n^2 c_{\mathbf{k}_*}c_\mathbf{y}}{\sqrt{s_{\mathbf{k}_*}s_\mathbf{y}}} \mathbf{k}_*^T (K-\sigma_n^2 I)^{-1}\mathbf{y}$, and hence 
\begin{equation}
\bar{f}_* = \frac{\sqrt{s_{\mathbf{k}_*}s_\mathbf{y}}}{\sigma_n^2 c_{\mathbf{k}_*} c_\mathbf{y}}\langle M \rangle.
\end{equation}
Note that since the outcome of a single run of the algorithm for measuring $M$ is a Bernoulli random variable, the process must be repeated a constant number of times to achieve a fixed variance on the estimate.
\item To approximate the variance $\mathbb{V}[{f}_*]$ the same procedure is followed as for the linear predictor, except that $\mathbf{v}$ is taken to be $\mathbf{k}_*$ instead of $\mathbf{y}$. This yields $\langle M\rangle = \frac{\sigma_n^2 c_{\mathbf{k}_*}c_\mathbf{y}}{s_{\mathbf{k}_*}} \mathbf{k}_*^T (K-\sigma_n^2 I)^{-1}\mathbf{k}_*$, and hence 
\begin{equation}
\mathbb{V}[{f}_*] = k(\textbf{x}_*,\textbf{x}_*)-\frac{s_{\mathbf{k}_*}}{\sigma_n^2 c_{\mathbf{k}_*}^2} \langle M \rangle.
\end{equation}
As with the linear predictor, $M$ is measured on multiple independent runs of the algorithm to yield the desired variance on the estimate.
\end{itemize}

Note that two variances $\text{Var}[\bar{f}_*]$ and $\mathbb{V}[{f}_*]$ are quantities differing in nature, the former arose as a by-product rooted in the uncertainty of quantum measurement when applying the quantum algorithm, while the latter is an inherent property of the regression. In order to estimate the variance and mean to within some desired precision $\delta$, it is necessary to repeat the measurement for sufficiently many times, such that $\text{Var}[\bar{f}_*]\leq \delta$ and $\text{Var}(\mathbb{V}[{f}_*]) \leq \delta$.

The above shows how QLA can be applied to computing two central objective quantities in Gaussian process regression problems. In the most ideal situation, where both $\mathbf{y}$ and $\mathbf{k}_*$ are sparse, and $(K+\sigma_n^2{I})$ is well-conditioned and highly-sparse, this procedure achieves an exponential speed-up over classical Gaussian process regression. Having the promising best-case scenario in mind, we now turn our attention to addressing factors which could limit the practicality of this procedure. 

The original QLA performs particularly well with sparse matrices. When the sparseness $s$ scales no faster than polylogarithmically in $n$, an exponential speed-up is possible. Fortunately, sparsely constructed Gaussian Processes are of significant interests in the context of a number of applications, particularly when the problem involves inference in large datasets \cite{Sparse2009}. For example, \cite{robotic} used sparse Gaussian process to construct a unified framework for robotic mapping. In the field of pattern recognition, sparse Gaussian processes have also been used to solve realistic action recognition problems \cite{recognition}. When $K$ is $s$-sparse, $\mathbf{k}_*$ will also be since it reflects the same dependencies as $K$. In such a case, even if $\mathbf{y}$ is dense, $\mathbf{y}$ can often be replaced by a sparse vector in the estimation of the linear predictor. This is because if one takes $T_x$ to be the matrix given by summing the first $x$ terms in the Taylor expansion of $(K-\sigma_n^2 I)^{-1}$, $\mathbf{k}_*^T T_x$ is $s^x$ sparse, and hence if $\mathbf{y'}$ is the $s^x$-sparse vector obtained from $\mathbf{y}$ by setting all entries to zero for which the corresponding entry in $\mathbf{k}_*^T T_x$ is zero $\mathbf{k}_* T_x \mathbf{y'} = \mathbf{k}_* T_x \mathbf{y}$. Even if $\mathbf{y}$ cannot be replaced by a sparse vector, the variance on estimation of the linear predictor will scale only linear in $n$, meaning that the estimation process must be repeated a linear number of times to reach constant error. This results in an algorithm which scales as $\tilde{O}(n)$. For more general applications where $s$ scales linearly with $n$, an exponential speed-up is not always guaranteed.

In order to apply QLA to Gaussian processes, the matrix $(K+\sigma_n^2{I})$ has to be well-conditioned. The ratio of largest and smallest eigenvalue $\kappa$ needs to stay low as $n$ increases in order for the matrix to be robustly invertible. In classical GPR, conditioning is already a well recognised issue. A general strategy to cope with the problem is to increase the noise variance $\sigma_n^2{I}$ manually by a certain amount to dilute the ratio $\kappa$ without severely affecting the statistical properties of the model. This increase in $\sigma_n^2{I}$ can be seen as a small amount of jitter in the input signal \cite{gaussian1998}. Therefore, for almost all practical purposes, we can assume the matrix is well-conditioned before applying QLA. Alternatively, when we have a sparse kernel where $s$ scales sufficiently slowly with $n$, we can employ the preconditioning technique described in \cite{precondi} which involves constructing a preconditoner matrix applied on the original system, and solve the well-conditioned modified linear system. As a result, conditioning does not provide a barrier to the use of QLA for GPR.

We have presented a novel procedure to apply the quantum algorithm for solving linear systems to Gaussian process regression modelling problems in supervised machine learning. By repeated sampling of the results of specific quantum measurements on the output states of QLA, the linear predictor and the associated variance can be estimated with bounded error with potentially exponential speed-up over classical algorithms. Our discussion of practicalities suggests that QLA is most advantageous when one is concerned with a sparse kernel and input vector, although the latter of these is not usually necessary. Under such circumstances, an exponential speed-up is achievable.

\textit{Acknowledgements}--- We thank Stephen Roberts for useful comments on the manuscript. JFF acknowledges support from the National Research Foundation and Ministry of Education Singapore. This material is based on research supported by the Singapore National Research Foundation under NRF Award No. NRF-NRFF2013-01.

\bibliographystyle{apsrev}
\bibliography{qgpr}
\end{document}